\documentclass[11pt]{article}
\pdfoutput=1
\usepackage{jheppub}

\usepackage{latexsym}
\usepackage{graphics}
\usepackage{amsmath}
\usepackage{amsfonts}

\newcommand{\lp}{\left(}
\newcommand{\rp}{\right)}

\usepackage{amssymb}
\usepackage{braket}
\usepackage{upgreek}

\newcommand{\lap} {\triangle}

\newcommand{\Lz}[1]{\lap_x #1 + \frac{a}{z} #1_z+ #1_{zz}}

\newcommand{\R}{\mathbb R}

\newcommand{\beq}{\begin{eqnarray}}
\newcommand{\eeq}{\end{eqnarray}}

\def\det{{\rm{det}}}

\usepackage{tikz}

\newtheorem{thm}{Theorem}[section]

\newtheorem{preremark}[thm]{Remark}

\numberwithin{equation}{section}

\title{Exact Form of Boundary Operators Dual to Interacting Bulk Scalar Fields in the AdS/CFT Correspondence}

\preprint{\today}

\author[a]{Gabriele La Nave}
\affiliation[a]{Department of Mathematics, University of Illinois, Urbana, Il. 61820}
\author[b]{Philip W. Phillips}
\affiliation[b]{Department of Physics and Institute for Condensed Matter Theory, University of Illinois, 1110 W. Green Street, Urbana, IL 61801}

\abstract{Using holographic renormalization coupled with the Caffarelli/Silvestre\cite{caffarelli} extension theorem, we calculate the precise form of the boundary operator dual to a bulk scalar field rather than just its average value. We show that even in the presence of interactions in the bulk, the boundary operator dual to a bulk scalar field is  an anti-local operator, namely the fractional Laplacian.  The  propagator associated with such operators is of the general power-law (fixed by the dimension of the scalar field) type indicative of the absence of particle-like excitations at the Wilson-Fisher fixed point or the phenomenological unparticle construction.   Holographic renormalization also allows us to show how radial quantization can be extended to such non-local conformal operators.}

\begin{document}

\maketitle

\section{Introduction}

In 2007, Caffarelli and Silvestre (CS)\cite{caffarelli} proved that standard second-order elliptic differential equations in the upper half-plane in ${\mathbb R}_+^{n+1}$ reduce to one with the fractional Laplacian, $\Delta^\gamma$, when one of the dimensions is eliminated to achieve ${\mathbb R}^n$.  For $\gamma=1/2$, the equation is non-degenerate and the well known reduction of the elliptic problem to that of Laplace's obtains. The precise statement of this highly influential theorem is as follows.  Let $f(x)$ be a smooth {\it bounded} function in ${\mathbb R}^n$ that we use to solve the extension problem
\begin{align}
g(x,0) &= f(x) \label{eq:dirichletboundary}\\
\Lz g &= 0 \label{eq:withy}
\end{align}
to yield a smooth {\it bounded} function, $g(x,z)$ in ${\mathbb R}^{n+1}_+$.
In these equations $f(x)$ functions as the Dirichlet boundary condition of $g(x,z)$ at the boundary $z=0$.  These equations can be recast in degenerate elliptic form,
\beq
{\rm div}(z^a\Delta g)=0\quad{ \in}\,{\mathbb R}_+^{n+1},
\eeq
which CS proved has the property that 
\begin{equation}\label{caff-limit}  \lim _{z\to 0^+} z^a \frac{\partial g }{\partial z} =C_{d,\gamma}\; {(-\lap)^\gamma f} \end{equation}
for some (explicit) constant $C_{n,\gamma}$ only depending on $n$ and $\gamma = \frac{1-a}{2}$
with $(-\lap)^\gamma$, the Reisz fractional Laplacian defined as
\beq\label{reisz}
(-\Delta_x)^\gamma f(x)=C_{n,\gamma}\int_{\R^n}\frac{f(x)-f(\xi)}{\mid {x-\xi}\mid^{n+2\gamma}}\;d\xi
\eeq
for some constant $C_{n, \gamma}$.
That is, the fractional Laplacian serves as a Dirichlet to Neumann map for elliptic differential equations when the number of dimensions is reduced by one.  Consider a simple solution in which, $g(x,0)=b$, a constant, but also $g_x=0$.   This implies  that $g(z)=b+z^{1-a}h$ with $(1-a)>0$.  Imposing that the solution be bounded as $z\rightarrow\infty$ requires that $h=0$ leading to a vanishing of the LHS of Eq. (\ref{caff-limit}).  The RHS  also vanishes because $-(\Delta_x)^\gamma b=0$.  As a final note on the theorem, from the definition of the fractional Laplacian, it is clear that it is a non-local operator in the sense that it requires knowledge of the function everywhere in space for it to be computed at a single point.  In fact, it is explicitly an anti-local operator.
Anti locality of an operator $\hat T$ in a space $V(x)$ means that for any function $f(x)$, the only solution to $f(x)=0$ ( for some $x\in V$ ) and $\hat T f(x)=0$ is $f(x)=0$ everywhere.  Fractional Laplacians naturally satisfy this property of anti-locality as can seen from their Fourier transform of Eq. (\ref{reisz}). 

Given that the AdS/CFT conjecture purports a fundamental relationship between bulk and boundary data on compact manifolds that differ by one spatial dimension, the CS extension theorem would then seem to be directly applicable to the boundary operator/bulk field correspondence.  However, until our recent \cite{gln2016} and related works on
scattering theory on conformally compact Einstein manifolds\cite{graham,Mazzeo}, the CS extension theorem has made no impact on the AdS/CFT conjecture.  This is unfortunate given the observation made by Chang and Gonzalez\cite{chang:2010} in 2010 that  for a compact manifold, $M={\mathbb R^n}$ with dimension $n$ and $X^{n+1}={\mathbb R}_+^{n+1}$ another compact manifold of dimension $n+1$ that has M as its boundary,  for the hyperbolic metric, $g_{\mathbb H}=\frac{dy^2+|dx|^2}{y^2}$,  the scattering operator for the equations of motion for a scalar field is solved by the fractional Laplacian using the Caffarelli-Silvestre extension theorem. Consequently, the bulk-boundary correspondence implicit in gauge-gravity duality can be made rigorous with the use of the Caffarelli-Silvestre extension theorem.  

It is this task that we undertake in this paper.  Building on  Chang/Gonzalez\cite{chang:2010} and our previous work\cite{gln2016}, we consider the scattering problem for an interacting massive scalar field in a spacetime that is asymptotically hyperbolic.  What we point out is that this problem has an exact solution and the boundary operator dual to the a bulk interacting scalar field is the fractional Laplacian identified by Caffarelli-Silvestre\cite{caffarelli} previously.  While our proof relies on the Caffarelli-Silvestre extension theorem, we also draw on the Wightman reconstruction theorem which indicates that the Hilbert space of states $\mathcal H$ and the operators of a quantum field theory are uniquely specified once the n-point functions are computed relative to some vacuum state $\Omega$.    

The Wightman reconstruction theorem is fundamentally related to the BDHM\cite{bdhm} formulation of the AdS/CFT correspondence.  Formally, the BDHM formulation of the AdS/CFT correspondence entails that  
\beq 
\label{correlators} \langle \mathcal O (x_1) \cdots \mathcal O (x_n)\rangle_{CFT} = \lim _{z\to 0} z^{-n\Delta} \langle \phi(x_1, z) \cdots \phi (x_n,z)\rangle_{bulk}.
\eeq
By Wightman's reconstruction theorem, the limit in Eq. \eqref{correlators} determines completely the form of the dual operator ${\mathcal O}$.  As long as the spacetime is asymptotically AdS\cite{hs}, the BDHM\cite{bdhm} and GKPW\cite{klebanov,klebanov2,Witten1998} formulations  are equivalent.  
In this note, we use the holographic renormalization technique of Heemskerk and Polchinski\cite{hp} along with the CS extension theorem\cite{caffarelli} to show that if the bulk is an interactive theory of the form,
\beq S_{bulk}= \int d^{d+1} x \sqrt{-g} \left( \sum _i |\nabla \phi _i|^2 + m_i^2 \phi _i^2 +\sum _{i,j} \lambda _{ij}\, \phi _i^2\phi _j\right)
 \eeq
 then, the boundary operators $\mathcal O$ behave to leading order as the fractional Laplacian, and they are thus {\it anti-local}, thereby extending our previous work\cite{gln2016} which applied strictly to the free case.   This operator identity removes the expectation value restriction in the AdS/CFT dictionary.   The significance of our result is that fractional Laplacians are not composite operators and hence should be viewed as the primary operators of the boundary theory. 
 
That the operator dual to a bulk scalar field should be an anti-local operator at the boundary is not unexpected because the boundary correlators\cite{Witten1998} all have anomalous dimensions.  If we consider the case of the Wilson-Fisher\cite{WF} fixed point where the correlation functions scale as
 \beq\label{unparticle}
 G(k,\omega)=k^{2-\eta}F(\frac{\omega}{k^z})
 \eeq
 where $z$ is the dynamical exponent and $\eta$, the anomalous dimension enters through the momentum dependence.  The same is true for any scale-invariant Green function, unparticles a case in point\cite{georgi}.  In both cases, anomalous powers of the momentum can be obtained from $\partial_\mu^\eta$, thereby suggesting that the operator content at the fixed point of WF and that underlying the unparticle construction should be thought of in terms of non-local operators.  The analogous reasoning applies to the AdS/CFT conjecture as well and it is this realization that unifies all such anomalous scaling as manifestations of boundary physics in a higher dimensional spacetime. 
 
 \section{Locality in QFT}

First a word on locality in QFT. Arguably the most crucial sense in which locality is imposed is the notion of {\bf micro-causality}, defined by the condition that for local operators $\mathcal O$ one has that 
 \beq\label{microcausality}
 [\mathcal O (x_1) , \mathcal O (x_2)] =0
 \eeq
 when $x_1$ and $x_2$ cannot be joined by a light-like geodesic.
This condition holds experimentally for the Standard Model (as verified at the L.H.C.) down to scales of order $10^{-20} m$.  Nothing we say violates micro-causality, as it can be seen by a straightforward application of Eq. \eqref{Oeq} and the fact that the fields $\varphi (x,t)$ are micro local.

The types of non-localities we are interested in here are non-localities of the operators in the sense defined in the previous section.  A {\bf local operator} is defined as a polynomial in fields $\phi$ and their (conventional) derivatives, or in mathematical terms they are (symbols of) differential operators in the fields $\phi$. 

The operators we derive here are (conformal versions of) fractional Laplacians in the fields, which cannot be constructed from any operator product expansion of local differential operators.  In this sense, the boundary operators we derive are not composite operators.  

 \section{Holographic Renormalization and dual fields}
 
  \subsection{Holographic Renormalization}
 
AdS/CFT is ultimately a correspondence between partition functions, where the bulk partition function is defined by means of  {\it boundary conditions}. Such boundary conditions are not given by a standard asymptotic condition of the field in the limit of approaching the conformal boundary, but rather in terms a renormalization procedure which is dictated by, for example, the classical solutions of the Gaussian theory (i.e., in Euclidean signature, solution to $-\Delta \phi+ m^2 \phi=0$) which have the asymptotic form
 \beq\label{asymptoticvalue}
\phi =F z ^{\frac{d}{2} -\gamma } + G z^{\frac{d}{2}+\gamma},\quad F,G\in\mathcal C^\infty(\mathbb H),\quad F=\phi_0+ O(z^2), \quad G= g_0 + O(z^2),
\eeq
where $\gamma=\frac{1}{2} \sqrt{ d^2+4m^2}$.
It is manifest from this asymptotic formula that there is no ambiguity as to what condition one should impose at the conformal infinity: $\frac{d}{2}+\gamma\equiv\Delta$ is always positive and therefore $G z^{\frac{d}{2}+\gamma}$ approaches $0$ towards the boundary.  Clearly the problematic term is $F z^{\frac{d}{2} -\gamma }$ when $\frac{d}{2} <\gamma$.  

One thus imposes that fields $\phi$ in the bulk be of the form $\phi = z^{\Delta_-} \phi _0$ as $z\to 0$ and $\phi _0$ is then named the boundary condition.  Here $\Delta_-=d/2-\gamma$.  For convenience in formulating the path integral in the partition function, we let $\beta(x)$ represent the non-normalizable mode with boundary condition, $\phi(x)=\epsilon^{-\Delta_-}\beta$ and $z=\epsilon$ will represent the inner cutoff for the integration of the bulk fields and $z=\ell$ the uppermost limit on the integration.
With this in mind, we follow the holographic renormalization formalism of Heemskerk and Polchinski \cite{hp} which is based on a separation of the partition function into UV and IR parts,
\beq \label{holographicrenorm}
Z_{bulk} [\beta]= \int \mathcal D \tilde \phi\; \Phi _{IR} [\tilde \phi, \ell] \,\Phi _{UV} [\beta, \tilde \phi;\epsilon, \ell],
\eeq
where
\beq \label{IRstate} \Phi _{IR} [\tilde \phi, \ell] =  \int \mathcal D \phi \mid _{z>\ell} e^{-S\mid _{z>\ell} }\eeq
and 
\beq \label{UVstate} \Phi _{UV} [\beta, \tilde \phi;\epsilon, \ell]= \int \mathcal D \phi \mid _{\epsilon<z<\ell} e^{-S\mid _{\epsilon<z<\ell} }.
\eeq
Here, given an interval $I$, the nomenclature $S\mid _{z\in I}$ means that we restrict the integration in the action with $z$ ranging only in $I$
and in Eq.( \eqref{holographicrenorm}) $ \phi$ (appearing in both equations \eqref{IRstate} and \eqref{UVstate}) satisfy the boundary conditions,
\beq
\label{boundarycond} \phi\mid _{z=\ell} = \tilde \phi \;\;\;\;\;\;  \phi\mid _{z=\epsilon} = \epsilon ^{\Delta _-} \phi _0.
\eeq
We can think of $\Phi _{IR} [\tilde \phi, \ell] $ (at least formally) as the state,
\beq
\Phi _{IR} [\tilde \phi, \ell] =e^{(\tau -\ell)H}| \tilde \phi\rangle
\eeq
and $\Phi _{UV} [\beta, \tilde \phi;\epsilon, \ell]$ as,
\beq
\Phi _{UV} [\beta, \tilde \phi;\epsilon, \ell]= \langle \tilde \phi |e^{(\ell -\epsilon)H}|  \epsilon ^{\Delta _-} \phi _0\rangle.
\eeq
In this formalism, one then takes the limit $\ell \to 0$ and $\epsilon\rightarrow 0$. It is manifest in this formulation that the part that is affected by the boundary condition is $ \Phi _{UV} [\beta, \tilde \phi;\epsilon, \ell]$.  Consequently, in computing an n-point correlation function, derivatives with respect to $\beta$ act only on $\Psi_{\rm UV}$, thereby pulling down factors of the bulk field, and as a consequence
\beq
\langle \mathcal O (x_1)\cdots \mathcal O(x_n)\rangle&=&\lim_{\ell\rightarrow 0}\int {\cal D}\tilde\phi\Psi_{\rm IR}\left[\frac{\delta }{\delta \beta (x_1)} \cdots \frac{\delta }{\delta \beta (x_n )}\psi_{\rm UV}\right]_{\beta=0}\nonumber\\&=&\lim _{z\to 0} z^{-n\Delta}\, \langle \tilde \phi(x_1,z) \cdots\tilde\phi(x_n,z)\rangle,
\eeq
generates the n-point function.  

As noted previously\cite{hs}, we write derivatives with respect to the radial coordinate as the limit of finite differences and as a consequence derivatives with respect to $\beta \epsilon^{d-\Delta}$ can be replaced with insertions of the conjugate momentum, $\partial_z\phi/\epsilon^{d-1}$.  Each such operation will generate a factor of $\tilde\phi$ and hence we arrive at the implicit definition of the dual boundary field (or dual operator)
\beq\label{Oeq}
{\cal O}(x)=\lim_{\epsilon\rightarrow 0}\epsilon^{1-\Delta}\partial_z(\phi(x,z)).
\eeq
As pointed out by Witten\cite{Witten1998}, both ${\cal O}$ and $\phi_0$ are conformal fields, differing only in their scaling dimensions.  Hence, there is no inherent disconnect between the left and right-hand sides of Eq. (\ref{Oeq}).  However, because states and operators are in a 1-1 correspondence via radial quantization, we can refer to ${\cal O}$ as the boundary dual operator as well, where the interpretation of the right-hand side (RHS) is strictly in terms of the states in the boundary Hilbert space defined in Eq. (\ref{IRstate}).  While the standard radial quantization is in terms of local operators, we generalize this notion below to apply to the type of non-localities we encounter here for a massive bulk scalar field.  Also, it is understood that the RHS has the non-renormalizable boundary term $(z^{\Delta_-}F)$ subtracted so that the limit in Eq. (\ref{Oeq}) is well defined. Although not explicitly written, it is implied that any subsequent expression of this kind has built-into it this subtraction procedure.  This expression is reminiscent of the more familiar expression viewed as the cornerstone of the AdS/CFT correspondence\cite{polchinski}
\beq\label{Oint}
{\cal O}(x)=\lim_{z\rightarrow 0}\epsilon^{-\Delta}\phi(x,z),
\eeq
which can be recast as Eq. (\ref{Oeq}) by inserting the conjugate momentum into  Eq. (\ref{Oint}). Eqs. (\ref{Oint}) and (\ref{Oeq}) are exact expressions not restricted to an average value with respect to some vacuum space $\Omega$.  

What we noticed previously is that if $\phi$ obeys the bulk equations of motion for a Klein-Gordon field, then the Caffarelli-Silvestre extension\cite{caffarelli} theorem can be used immediately to take the limit in Eq. (\ref{Oeq}) and obtain the formal expression for ${\cal O}$.  For completeness, we review this theorem here.  To set up their result, we note that substituting Eq. (\ref{asymptoticvalue}) into Eq. (\ref{Oeq}) yields
\beq\label{ofinal}
{\cal O}(x)=\lim_{z\rightarrow 0}G=g_0.
\eeq
As a consequence, determining ${\cal O}(x)$ simply requires that we identify $g_0$.   This is where the Caffarelli-Silvestre extension theorem can be put to use.  The relationship between our problem and theirs is given that $\phi$ solves the equations of motion for the bulk scalar field,
the function
\beq\label{geq}
 g=z^{\gamma -\frac{d}{2}}\, \phi=F+z^{2\gamma}G
 \eeq
 solves the Caffarelli-Silvestri extension problem.  Note
 \beq\label{gfinal}
 \lim_{z\rightarrow 0} g(x,z)=\lim_{z\rightarrow 0}( F+z^{2\gamma}G)=\phi_0
 \eeq
 from the asymptotic expansion for $F$ and $G$ in Eq. (\ref{asymptoticvalue}).  So in Caffarelli/Silvestre $\phi_0$ plays the role of $f(x)$.  Substitution of Eq. (\ref{geq}) into the LHS of Eq. (\ref{caff-limit}) leads to
 \beq
 \lim _{z\to 0^+} z^a \frac{\partial g }{\partial z}&=&\lim _{z\to 0^+} z^{1-2\gamma} \frac{\partial}{\partial z} (F+z^{2\gamma}G )\nonumber\\
&=&2\gamma g_0,
\eeq
where we have used the asymptotic expansion for the coefficients $F$ and $G$ in Eq. (\ref{asymptoticvalue}).  This equation explicitly defines ${\cal O}$ by Eq. (\ref{ofinal}).   Consequently, using the RHS from the Caffarelli/Silvestre extension theorem, Eq. (\ref{caff-limit}),  ${\cal O}$ is given by ${\cal O}(x)=C_{\cal O}(-\Delta)^\gamma  \phi _0$; that is,
 \beq\label{Gexp}
 G= C_{\cal O}(-\Delta)^\gamma  \phi _0+O(z^2).
 \eeq
 While it is standard in the AdS/CFT correspondence\cite{klebanov,klebanov2,bdhm} to view the coefficient $G$ as the expectation value (relative to some vacuum state) of the dual operator ${\cal O}$, this interpretation is only natural within an on-shell evaluation of the action.   What we show is that the form obtained here for $G$ is valid even beyond the on-shell approximation and hence our evaluation of $G$ is not  restricted to a particular choice of the vacuum.  Hence, by using the Caffarelli/Silvestre extension theorem  and the weak nature of the interactions at the boundary,  $G$ determines ${\cal O}$ to first order. 
 
 As a consequence, the proper interpretation of Eq. (\ref{Gexp}) is in terms of the  $\Phi_{\rm UV}$ states defined earlier. Because states are operator-valued distributions in a Hilbert space and operators and states are equivalent in a CFT, both sides of the expression defining ${\cal O}$ are of the same type.    To deconstruct this expression, we rely on the work of Chang and Gonzalez\cite{chang:2010} who showed, based on the Caffarelli/Silvestre extension theorem, that the coefficients $F$ and $G$ are in general related via a Dirichlet to Neumann map and the map is provided by the  fractional Laplacian.  Since $F$ to leading order is $\phi_0$, then Eq. (\ref{Gexp}) must be true as we have shown by evaluating Eq. (\ref{Oint}) exactly.  An alternative way of looking at this expression is that the functional derivative of $\delta\langle{\cal O}\rangle/\delta\phi_0(k)$  where $\langle\cdots\rangle$ refers to expectation value over some vacuum state that is independent of the source fields $(\phi_0$), that is, it is determined by the excitations of the Hilbert space, yields the Green function.  It is easy to verify that appropriate derivatives of the above expression yields $k^{2\gamma}$, the correct result for a scalar field.    Making contact with the Wilson-Fisher fixed point discussed in the introduction, such anomalous momentum dependence can only arise if the corresponding operator in real space is non-local.

\subsection{Dual Field}

To summarize, in the previous section we used the Caffarelli/Silvestre extension theorem\cite{caffarelli} to show that in Gaussian Klein-Gordon theory (i.e., $S_{bulk}(\phi) =\int \, \sqrt{-g}\,d^{d}xdz\left( |\nabla \phi |^2+m^2\phi^2\right)$) the field $\mathcal O$ dual to $\phi _0$, 
which is determined by
\beq\label{limiteq}
\mathcal O(x) = \lim _{z\to 0} z^{1-\Delta } \partial _z\phi(x,z)
\eeq
is in fact equal to the conformal fractional Laplacian $(-\Delta )^\gamma \phi_0$ (here $\phi_0$ i the boundary condition), when the mass squared $m^2$ is tachyonic: $-\frac{d^2}{4}\leq m^2 <0$. What we want to show now is that even in the presence of interactions, the same result holds.

We now turn on interactions and consider
\beq \label{S}
S_{bulk}(\phi) =\int \, \sqrt{-g}\,d^{d}xdz\left( |\nabla \phi |^2+m^2\phi^2+V(\phi)\right).
\eeq
As observed by Harlow and Stanford (\cite{hs}), we simply need to evaluate the UV part (as it is the one influenced by the boundary conditions).
%
%
We then consider
\beq \label{UV}
\int _{\phi(\epsilon)=\epsilon ^{\Delta _-} \beta}^{\phi (\ell) =\tilde \phi} 
\mathcal D \phi \mid _{\epsilon<z<\ell} e^{-S_{bulk} [\phi]\mid _{\epsilon<z<\ell} }
\eeq
with
\beq \label{SUV}
S_{bulk} [\phi]\mid _{\epsilon<z<\ell} =\int \, d^{d}x\int_{\epsilon}^\ell \, \frac{dz}{z^{d+1}} \left( \frac{z^2}{2} \left((\partial _z \phi)^2+ |\nabla _x\phi |^2\right) +V(\phi)\right).
\eeq
Following \cite{hs}, we make the rescaling $y=\frac{z}{\ell}$ so that the integral in Eq. \eqref{SUV} transforms to
\beq \label{SUV2}
S_{bulk} [\phi]\mid _{\epsilon/\ell<y<1} =\frac{1}{\ell^d}\int \, d^{d}x\int_{\epsilon/\ell}^1 \, \frac{dy}{y^{d+1}} \left( \frac{y^2}{2} (\partial _y \phi)^2+ \frac{y^2\ell^2}{2} |\nabla _x\phi |^2+V(\phi)\right),
\eeq
and now the prefactor, $\frac{1}{\ell^d}$, renders the  saddle point approximation exact.  The effective equations of motion are 
\beq
\left\{ \begin{aligned} &y^2 \partial _y^2 \phi  +(1-d) \, y\, \partial _y \phi + \frac{y^2\ell^2}{2} \Delta _x\phi = V'(\phi)\\& \phi(1,x)=\tilde \phi, \;\;\;\; \phi(\frac{\epsilon}{\ell}, x)= \phi _0 \end{aligned}\right.,
\eeq
where the top equation is  the Laplacian for the rescaling of the hyperbolic metric.  Even in the presence of the $ \Delta _x\phi $ term, the classical solutions have the asymptotics (see the appendix for a quick sketch of an argument),

 \beq\label{asymptoticvalue-interaction}
\phi =F y^{\frac{d}{2} -\gamma } + G y^{\frac{d}{2}+\gamma},\quad F,G\in\mathcal C^\infty(\mathbb H),\quad F=\phi_0+ O(y^2), \quad G= (-\Delta \phi _0)^\gamma + O(y^2),
\eeq
we identified earlier when $\gamma <\frac{d}{2}$ (i.e. $m^2<0$) .  Hence, the fractional Laplacian appears explicitly as the boundary operator dual to the bulk scalar field.  Although it is the $ \Delta _x\phi $ term that accounts explicitly for deviations around the 1-d classical path, retaining it does not change the asymptotics.  Hence, while  this term vanishes in the scaling limit, this is not a prerequisite for the identification of the dual boundary operator. 

\section{Radial quantization of conformal (non-local) operators}

\subsection{The conformal Laplacian}\label{conformalS}
Recall that for an asymptotically $d+1$ AdS space-time  $(M, d\tau ^2=g^+ _{\mu \nu}\, dx^\mu \otimes dx^\nu)$ with conformal boundary $X$, one 
defines the scattering operator as follows.
 Solutions to
\begin{equation}\label{equation-GZ}
-\Delta_{g} u-s(n-s)u=0,\quad\mbox{in } X
\end{equation}
have the form
\begin{equation}\label{general-solution}u = F \rho ^{n-s} + H\rho^s,\quad F,H\in\mathcal C^\infty(X),\quad F|_{\rho=0}=f,\end{equation}
for all $s\in\mathbb C$ unless $s(n-s)$ belongs to the pure point spectrum of $-\Delta_{g}$.
The {\it scattering operator} on $M$ is defined as
 $$S(s)f = H|_X.$$ 

The conformally covariant fractional powers\cite{chang:2010} of the Laplacian (on the conformal boundary) is determined by
 \begin{equation}\label{P-operator} P_\gamma[d\tau ^2, h] := D_\gamma \; S\lp\frac{d+1}{2}+\gamma\rp,\quad D_\gamma=2^{2\gamma}\frac{\Gamma(\gamma)}{\Gamma(-\gamma)},\end{equation}
for $s=\frac{d}{2}+\gamma$, $\gamma\in \lp 0,\frac{d}{2}\rp$, $\gamma\not\in \mathbb N$. 
For simplicity of notation, since the space-time metric $d\tau ^2$ will be fixed, we simply write $ P_\gamma[h] $ for $ P_\gamma[d\tau ^2, h] $.
Then an important property of the conformal fractional Laplacian is that when we change $h$ in the conformal class $[h]$ by $h_u=e^{2u}h$,
\beq \label{fractionalLapconform}
P_\gamma[h_u] (\phi)= e^{-(w+\gamma)u} P_\gamma[h] \left(e^{(w)u} \phi\right),
\eeq
where $w=\frac{d}{2}-\gamma$. 

A choice of a Lorentzian metric on a manifold $M$ of dimension $d+1$ is equivalent to a choice of an {\it orthonormal } frame bundle of $T^*M$. Choosing a conformal class is equivalent to a reduction of the structure group.
For any real number $\alpha\in \mathbb R$, one obtains a ~1-dimensional irreducible representations of $SO(1,d)$ given by $\det (A) ^{\frac{\alpha}{d+1}}$ which gives rise to a line bundle $L_\alpha$ for a fixed conformal structure. Choosing a metric $g$ in the conformal class $[g]$ is tantamount to choosing a trivialization $\tau _{g,\alpha} : L^\alpha \to M\times \mathbb R$ and changing $g$ by $e^{2w}\, g$ has the effect of modifying the trivialization to  $e^{-\alpha w}\tau_{g,\alpha}$. 
Considering $\hat P_\gamma[h] = \tau _{g, \frac{d}{2} -\gamma}^{-1}  \circ P_\gamma[h] \circ  \tau _{g, \frac{d}{2} +\gamma}$ gives rise to an operator
\beq \label{conformalfract} \hat P_\gamma[h]  : L^{\frac{d}{2} -\gamma} \to L^{\frac{d}{2} +\gamma},
 \eeq
which is a {\it conformally invariant} non-local operator. 

Also, using the formula in Eq. \eqref{fractionalLapconform},  one obtains a {\it conformally invariant non-local action} by considering
\beq\label{comnfinvaaction}
S_\gamma [\phi]= \int _M \;dV_h\; \phi\, P_\gamma[h](\phi)\,
\eeq
where $\phi$ is a section of $L^{\frac{d}{2}-\gamma}$ and where $dV_h$ is the volume form of the metric $h$.  Hence, the boundary dual to a scalar field can be formulated strictly conformally as well.

\subsection{Conformal operators and radial quantization}

Witten's association of ${\cal O}$ with the limiting form of the field $\phi_0$ was established by integrating the action by parts.  We have used an alternative equivalent approach here based on Eq. (\ref{Oeq}).  However, the conclusion that ${\cal O}$ is antilocal still permits a formulation in terms of radial quantization that is typically performed for conformal local operators.  This operator acts on a Hilbert space modeled on the space of sections of powers of the line bundle $L$ as in the previous section (rather than functions). Let's call our Hilbert space of states $H$.

Recall that the classical construction produces a state out of a "localized"{\begin{footnote} { Here we opt for the nomenclature "localized" as opposed to standard one "local", simply because we want to argue that one can localize at a point even a conformal, not necessarily local operator  } \end{footnote}} operator by the path integral method as follows.  Hence, the current extension is non-trivial.  We focus on the ~2-dimensional case for simplicity. One transposes a "localized" operator $\Phi$ along the radial direction by integrating over the annulus $A=\{ z\in \mathbb C: r<|z|<R\}$ using
\beq
\Phi(\phi _f, r)= \int \,\mathcal D \phi _i (x) \, \int_{\phi (x, r) = \phi_i} ^{\phi(x,R)= \phi _f} \mathcal D \phi \, e^{\frac{i}{\bar h}\int_r^R \, d\rho\, L}\Phi (\phi _i,r)
\eeq
which is in general an operator from $H$ (viewed as the states on the inner circle) to $H$ (viewed as the sates on the outer circle)
and then one lets $r\to 0$ to then obtain
\beq
\begin{aligned} \Phi(\phi _f, 0)&= \int \,\mathcal D \phi _i  \, \int_{\phi ( 0) = \phi_i} ^{\phi(x,R)= \phi _f} \mathcal D \phi  \,e^{\frac{i}{\bar h}\int_0^R \, d\rho\, L}\Phi (\phi _i,0)\\& =\int ^{\phi(x,R)= \phi _f} \mathcal D \phi\, e^{S(\phi)}\Phi (\phi _i,0) \end{aligned}
\eeq
which produces an operator from the space of localized operators (i.e operators of the form $\Phi (\phi _i,0)$) to $H$. 
It is now evident, from the discussion in section \ref{conformalS} that this construction generalizes to the case of conformal operators which are not necessarily local. This is because the limit for $r\to 0$ is a rescaling limit, and conformal operators are in particular scale invariant.  Hence, in spacetimes which are asymptotically AdS, the radial quantization of non-local operators does not pose a problem.

\section{Final Remarks}

We have shown using the Caffarelli/Silvestre\cite{caffarelli} extension theorem that the precise form of the boundary operator dual to a bulk scalar field can be obtained in general.  Our construction relies on the Heemskerk/Polchinski\cite{hp} holographic renormalization scheme which permits a precise formulation of the boundary operator.  Within this procedure, the radial quantization of non-local operators can be carried over because the scale invariance in the scaling limit $\ell\rightarrow 0$.  The fact that the boundary operator is non-local is not unexpected given that the boundary correlation functions possess anomalous dimensions.  Given that even a simple scalar field generates boundary non-locality, our work raises the question as to which kinds of fields in the bulk actually have local boundary duals.  As of this writing and coupled with the Cafferelli/Silvestre\cite{caffarelli} extension theorem which demonstrates that the limiting form of boundary differential operators necessarily involves the fractional Laplacian, this question remains open.  

Our work establishes an obvious connection between the unparticle construction of Georgi\cite{georgi}
and the 2-point correlators at the boundary of AdS.  Because the anomalous dimension at the boundary is simply related to the mass of the bulk field, anomalous dimensions in the unparticle Green functions can all be engineered within the AdS/CFT procedure simply by tuning the mass of the bulk scalar field. Hence, in terms of the operator content at the boundary, unparticles should just be thought of simply as a realization of the AdS/CFT correspondence.  In fact, Domokos and Gabadadze\cite{domokos} have recently shown that unparticles give rise to non-local actions involving fractional powers of the Laplacian precisely of the kind found here.  Consequently, the current work unifies unparticles and the larger AdS/CFT correspondence.  

\section*{Acknowledgements}
We thank the NSF DMR-1461952 for partial funding of this project .

\section{Appendix: Asymptotics}
Here we intend to show the statement that classical solutions of:
\beq
\left\{ \begin{aligned} &\Delta _ {\mathbb H,z}\phi= V'(\phi)\\& \phi(1,x)=\tilde \phi, \;\;\;\; \phi(\frac{\epsilon}{\ell}, x)= \phi _0 \end{aligned}\right.
\eeq
with $\Delta _{\mathbb H,z}$ indicating the hyperbolic Laplacian, have the asymptotics,
\beq\label{asymptoticvalue-interaction}
\phi =F z^{\frac{d}{2} -\gamma } + G z^{\frac{d}{2}+\gamma},\quad F,G\in\mathcal C^\infty(\mathbb H),\quad F=\phi_0+ O(z^2), \quad G= (-\Delta)^\gamma \phi _0 + O(z^2),
\eeq
Under the simplifying assumption that $V$ is analytic and that $V=\frac{m^2}{2} \phi ^2 + R(\phi)$ where $R(\phi)$ only has terms higher than third order, this is done by assuming the ansatz:
$$\phi= z^{\Delta _-} \sum _{k=1}^{\infty} a_k(x)z^k + z^{\Delta _+} \sum _{k=1}^{\infty} b_k(x)z^k$$
In the relevant case in which we assume $\phi$ to only depend on $z$, the equation reduces to:
$$z^2 \phi ''+ (1-d) z \phi'- m^2 \phi= R'(\phi)$$
then one readily sees the lowest order term of $G= \sum _{k=1}^{\infty} b_k(x)z^k$ is the one that induces the fractional Laplacian.  Since the hyperbolic Laplacian explicitly contains the term $ \Delta _x\phi$, we see that the identification of the fractional Laplacian as the boundary operator dual to the bulk
interacting field does not necessitate the 1-dimensional saddle-point approximation.

\providecommand{\href}[2]{#2}\begingroup\raggedright\endgroup

\end{document}